# Visible-light photocatalytic oxygen production on a high-entropy oxide by multiple-heterojunction introduction

Parisa Edalati[a], Yuta Itagoe[b], Hironori Ishihara[c], Tatsumi Ishihara[b,d], Hoda Emami[e], Makoto Arita[f], Masayoshi Fuji*[,a,g] and Kaveh Edalati*[,d]

[a] Department of Life Science and Applied Chemistry, Nagoya Institute of Technology, Tajimi 507-0071, Japan
[b] Department of Applied Chemistry, Faculty of Engineering, Kyushu University, Fukuoka 819-0395, Japan
[c] School of Science, Fukuoka university, Fukuoka 814-0133, Japan
[d] WPI, International Institute for Carbon-Neutral Energy Research (WPI-I2CNER), Kyushu University, Fukuoka 819-0395, Japan
[e] Corem, 1180 Rue de la Mineralogie, Quebec, G1N 1X7, Canada
[f] Department of Materials Science and Engineering, Faculty of Engineering, Kyushu University, Fukuoka 819-0395, Japan
[g] Advanced Ceramics Research Center, Nagoya Institute of Technology, Tajimi 507-0071, Japan

**Abstract**
High-entropy oxides (HEOs), as multi-component ceramics with high configurational entropy, have been of recent interest due to their attractive properties including photocatalytic activity for $H_2$ production and $CO_2$ conversion. However, the photocatalytic activity of HEOs is still limited to ultraviolet light. In this study, to achieve visible-light-driven photocatalysis, 10 different heterojunctions were simultaneously introduced in the Ti-Zr-Nb-Ta-W-O system. The oxide, which was synthesized by a high-pressure torsion method and oxidation, successfully produced oxygen from water under visible light without co-catalyst addition. The photocatalytic performance was attributed to high visible-light absorption, narrow bandgap, appropriate band structure, presence of multiple heterojunctions and accordingly easy electron-hole separation and slow recombination. These results not only show the potential of high-entropy oxides as new visible-light-active photocatalysts, but also introduce the multiple-heterojunction introduction as a strategy to achieve photocatalysis under visible light.

* Corresponding Authors
  Kaveh Edalati (E-mail: kaveh.edalati@kyudai.jp; Tel: +81-92-802-6744)
  Masayoshi Fuji (E-mail: fuji@ fuji@nitech.ac.jp; Tel: +81-57-227-6811)

# 1. Introduction

Photocatalysis is a chemical process under the light in the presence of a light-absorbing catalyst, which is known as a photocatalyst [1-4]. Since the process only uses solar energy, it is considered a green chemical pathway to produce hydrogen from water, convert $CO_2$ to useful hydrocarbons, or decompose toxic compounds [5-7]. Photocatalytic oxygen evolution reaction from water splitting is another chemical reaction that has received significant attention in the photocatalyst field [5]. In the photocatalytic water splitting process, electrons in the valence band of the semiconductor absorb the solar light and transfer to the conduction band, and the transferred electrons take part in the reduction reaction of water to produce hydrogen, and the holes in the valence band take part in the oxidation reaction to produce oxygen [8,9]. Since the photocatalytic oxygen evolution reaction needs four holes, it is more difficult than many other photocatalytic reactions including hydrogen production [9,10].

The current challenge in photocatalytic water splitting is finding an appropriate catalyst that can act under visible light with high efficiency. Such ideal photocatalysts should have several main features such as a narrow bandgap, easy electron and hole separation, a low recombination rate of electrons and holes, and an appropriate electronic band structure to cover water splitting reactions [10]. However, most photocatalysts only act under ultraviolet (UV) light due to their wide bandgap. There are several typical attempts for achieving photocatalytic water splitting activity under visible light which contain impurity doping [11], defect engineering [12], surface modification [13], introducing mesoporous structures [14], nanosheet production [15], high-pressure phase generation [16] and introducing heterojunctions [17].

Heterojunctions are formed by making heterostructured composites of at least two phases or compounds, which have high light absorbance and appropriate band structure at the interface [17,18] There are different types of heterojunctions, but the most popular one can be described here as a dual-phase system [18]. In this type, the electrons are excited from the valence band to the conduction band in both phases by absorbing light photons with energies higher than their bandgaps. Following this excitation, the electrons transfer from the conduction band of one material (with a higher bottom of conduction band energy) to the conduction band of another material, and the holes transfer from the valence band of one material (with a lower top of valence band energy) to the valence band of other material, and this leads to an extended electron-hole separation and migration [18]. There are several successful reports on designing heterojunctions to achieve visible-light photocatalytic water splitting such as α-$Fe_2O_3$-Nanorod/Graphene/$BiV_{1-x}Mo_xO_4$ [19], g-$C_3N_4$/$Ag_3PO_4$ [20], $CaIn_2S_4$/g-$C_3N_4$ [21], NiO/Ni/$TiO_2$ [22], $Bi_2MoO_6$/$Bi_2Mo_3O_{12}$ [23] and g-$C_3N_4$/$TiO_2$ [24] In all these successful examples, the presence of one chemically stable material with high light absorbance in the visible-light region is essential.

High-entropy ceramics including high-entropy oxides (HEOs) and oxynitrides (HEONs) were recently introduced as stable catalysts with large light absorbance in the visible-light region [25-27], but there has been no success to achieve visible-light-driven photocatalysis on these materials. High-entropy ceramics are described as materials containing at least five cations, having a configurational entropy higher than $1.5R$ ($R$: gas constant) and accordingly a low Gibbs free energy and high stability [28-30]. Among high-entropy ceramics, HEOs are the most popular ones which are used for various applications including Li-ion batteries [31-33], magnetic components [34-36], dielectric components [37-39] and thermomechanical applications [40-42]. Moreover, there are some reports regarding the successful application of HEOs as stable catalysts for catalytic CO oxidation [43], catalytic combustion reaction [44], catalysis in electrochemical capacitors [45], electrocatalytic oxygen evolution [46], electrocatalysis in metal-air batteries [47], electrocatalytic hydrogen generation [25] and photocatalytic carbon dioxide conversion [26,48]. In addition to high entropy and resultant high chemical stability for catalysis, the existence of inherent lattice strain and defects such as oxygen vacancies caused by the presence of various elements with different atomic radii makes high-entropy ceramics attractive for photocatalysis [49]. Defects and strained regions can facilitate trapping the electrons and act as active sites to absorb the reactant for photocatalytic reactions and improve the catalytic activity [15,49].

Motivated by earlier studies on photocatalysts, heterojunctions and HEOs, one can expect that a combination of HEOs and heterojunctions can be an effective strategy to achieve visible-light-driven photocatalysis for difficult reactions such as oxygen evolution.

In this work, a high-entropy oxide in the Ti-Zr-Nb-Ta-W-O system with multiple heterojunctions was designed and synthesized for visible-light-driven photocatalytic oxygen production. To design the material, the elements which can produce oxides with the $d^0$ oxidation states were selected. The first selected combination of Ti, Zr, Hf, Nb and Ta led to the formation of a dual-phase HEO with a rather wide bandgap and UV-light-driven photocatalytic activity. Statistically, it is more desirable to have multiple heterojunctions for a system with unknown band structures, so that at least one of the heterojunctions can satisfy the requirements for visible-light-driven photocatalysis. Therefore, a second material was prepared by a combination of Ti, Zr, Nb, Ta and W which led to the formation of five phases and 10 heterojunctions. The co-presence of multiple heterojunctions together with an overall low bandgap of the high-entropy oxide resulted in the oxygen production from water under visible light.

## 2. Experimental procedures
### 2.1. Materials and synthesis

Although many different methods were developed in recent years for the synthesis of HEOs [28-47], a three-step synthesis was used in this study. First, small pieces of high purity Ti (99.9%), Zr (99.5%), Nb (99.9%), Ta (99.9%) and W (99.9%) in equal atomic fractions were arc-melted under an argon atmosphere in a water-cooled copper crucible to produce a metallic ingot with a total mass of ~12 g. To improve the chemical homogeneity, the ingot was rotated and remelted 20 times in the arc melting furnace. After arc melting, disc samples with 10 mm diameter and 0.8 mm thickness were prepared from the ingot using an electric discharge machine. The metallic discs were further homogenized by a high-pressure torsion (HPT) method [50], schematically shown in Fig. 1(a). This high-pressure method was selected due to its reported capability in synthesizing homogenous high-entropy alloys [51,52]. The HPT process was conducted at room temperature under a pressure of 6 GPa with a rotation speed of 1 rpm for 100 turns. Finally, the HPT-processed discs were crushed and oxidized by heating in a hot air atmosphere at a temperature of 1373 K for 40 h. The oxidized material was in the form of an orange powder, shown in Fig. 1(b), containing particles with various morphologies with an average size of 470 nm, as shown in Fig. 1(c). The evolution of material during each step of synthesis is shown in the X-ray diffraction (XRD) profiles in Fig. 2: the arc-melted alloy had a dual-phase structure containing two body-centered cubic (BCC) structures (with lattice parameters of $a = 0.326$ nm and 0.343 nm), it transformed to a single-phase high-entropy alloy with the BCC structure ($a = 0.330$ nm) after HPT processing, and it finally transformed to an oxide with five different phases after oxidation. Comparing the mass of the sample before and after oxidation suggested that the high-entropy oxide should have a general composition of TiZrNbTaWO$_{12}$.

### 2.2. Characterization
The oxide was characterized by various techniques, as described below.

(i) The crystal structure and phase identification were examined at room temperature by the XRD method using Cu Kα radiation. The diffraction patterns were analyzed by the Rietveld method using the Fullprof software [53]. Crystal structures were visualized by using the VESTA software. Moreover, micro-Raman spectroscopy using a 532 nm laser was carried out to investigate the crystal structure homogeneity.

(ii) The microstructural features and distribution of elements were examined using scanning electron microscopy (SEM) equipped with an energy-dispersive X-ray spectroscopy (EDS) detector under an acceleration voltage of 15 keV. Nanostructural features were examined by aberration-corrected transmission electron microscopy (TEM) using the selected area electron diffraction (SAED), bright-field (BF) images, dark-field (DF) images, high-resolution images and fast Fourier transform (FFT) analysis under an acceleration voltage 200 keV. The distribution of elements at the nanometer scale was examined by scanning-transmission electron microscopy (STEM) using the high-angle annular dark-field (HAADF) images and EDS analysis under an acceleration voltage of 200 keV. The samples for SEM and TEM were prepared by dispersing the oxide powder on a copper stage and a carbon gride, respectively.

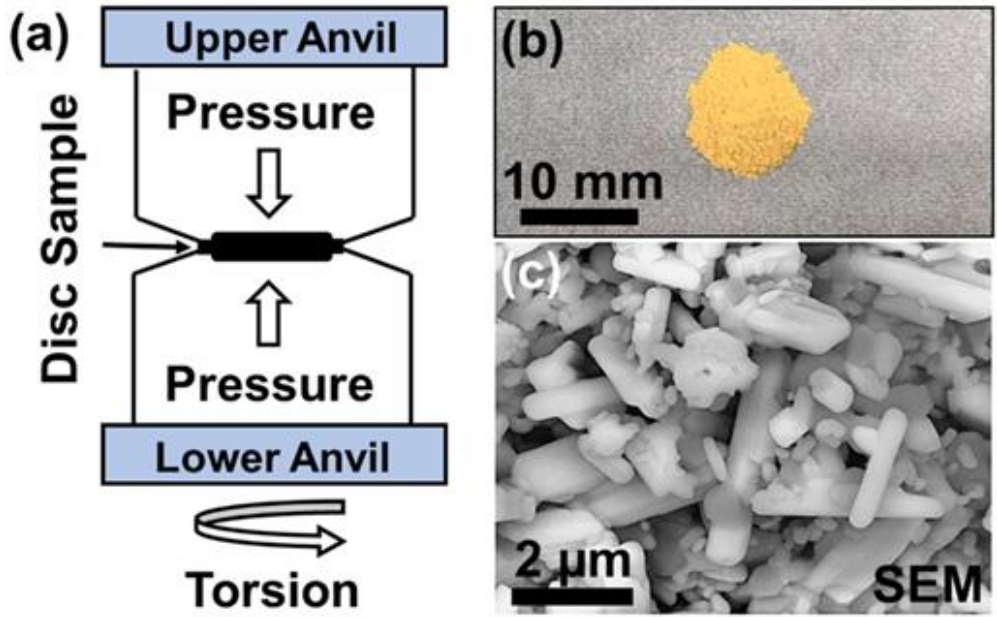

Fig. 1 (a) Schematic illustration of HPT method. (b) Appearance and color of HEO. (c) SEM image of HEO.

(iii) To investigate the size distribution of particles in aqueous suspension, in addition to SEM observation, the dynamic light scattering (DLS) method was performed using a Zetasizer Nano-S facility equipped with a 4 mW He-Ne laser (633 nm) and 173 diffraction angle.

(iv) The oxidation states of elements were examined by X-ray photoelectron spectroscopy (XPS) using Al Kα radiation with a wavelength of $\lambda = 0.989$ nm. Peak deconvolution was conducted on XPS profiles by considering the standard relations of intensity ($I$) and binding energy ($E$): $I(f_{7/2}) : I(f_{5/2}) = 4:3$, $I(d_{5/2}) : I(d_{3/2}) = 3:2$, $I(p_{3/2}) : I(p_{1/2}) = 2:1$, $E(\text{Ti } 2p_{1/2}) - E(\text{Ti } 2p_{3/2}) =$

5.54 eV, $E$(Zr 3d$_{3/2}$) - $E$(Zr 3d$_{5/2}$) = 2.43 eV, $E$(W 4f$_{5/2}$) - $E$(W 4f$_{7/2}$) = 1.71 eV, $E$(Nb 3d$_{3/2}$) - $E$(Nb 3d$_{5/2}$) = 2.72 eV, $E$(Ta 4f$_{5/2}$) - $E$(Ta 4f$_{7/2}$) = 1.91 eV [54].
(v) The light absorbance was examined using UV-vis diffuse reflectance spectroscopy and the bandgap was estimated using the Kubelka-Munk analysis. The top of the valence band was determined using ultraviolet photoelectron spectroscopy (UPS) using a He-I UV light source and a DC bias of -4 V. The bottom of the conduction band was calculated by subtracting the bandgap from the top of the valence band.
(vi) To examine the electron-hole recombination, photoluminescence (PL) spectroscopy was recorded using a UV laser source with a wavelength of 325 nm.
(vii) The specific surface area of oxide, which is of importance for photocatalysis, was determined by nitrogen gas adsorption and using the Brunauer-Emmett-Teller (BET) method.

*2.3. Photocatalysis*

A photocatalytic test was conducted for oxygen production from water under visible light irradiation. For this test, 30 mg of sample was dispersed in 30 mL of AgNO$_3$ (20 mM), in which AgNO$_3$ was used as a sacrificial agent. After stirring in dark for 1 h and confirming the absence of oxygen and hydrogen gasses, the suspension was irradiated by light. The light source was a 300 W Xe lamp (PE300BUV, Perkin Elmer) equipped with a 420 nm cut-off filter and the light intensity on the suspension was 0.968 W/cm$^2$. The solution was stirred continuously during the irradiation and the amount of oxygen and hydrogen were quantified using a gas chromatograph (GC-8A, Shimadzu) and integrator (C-R6A Chromatopac, Shimadzu). The stability of the catalyst was determined by XRD, TEM, XPS and UV-vis analyses after photocatalysis. It should be noted that a blank test was also conducted without the addition of catalyst to the liquid to be sure about the absence of oxygen from other sources during irradiation.

## 3. Results
*3.1. State of elements and crystal structure*

The crystal structure of oxide is shown in Fig. 2(c) using XRD profiles. Fig. 2(c) demonstrates that five different phases including one orthorhombic phase, two monoclinic phases, and two tetragonal phases are formed after the oxidation of HPT-processed material. The Rietveld analysis suggests that the weight fraction of orthorhombic (Ima2), monoclinic-I (C12/m1), monoclinic-II (P12/m1), tetragonal-I (P4/nmm) and tetragonal-II (P42/mnm) phases are 29.4, 38.6, 26.4, 3.9 and 1.7 wt%, respectively. Crystallographic information of these phases including their space groups, lattice parameters, lattice angles, their fractions and their simulated crystal structure visualized using the VESTA software are presented in Table. 1. Fig. 3 shows the micro-Raman spectra measured in seven different positions of the oxidized sample. In all these seven positions, similar Raman spectra with major wavenumbers at 120, 240, 430, 680 and 1000 cm$^{-1}$ are observed. This indicates that the distribution of phases is well uniform at the submicrometer level so that they cannot be differentiated using the micro-Raman spectroscopy. Such uniform distribution of phases suggests that all phases can possibly have joint interfaces as heterojunctions. If it is assumed that every two phases can produce one heterojunction [17,18], the presence of five phases can lead to the formation of 10 possible heterojunctions.

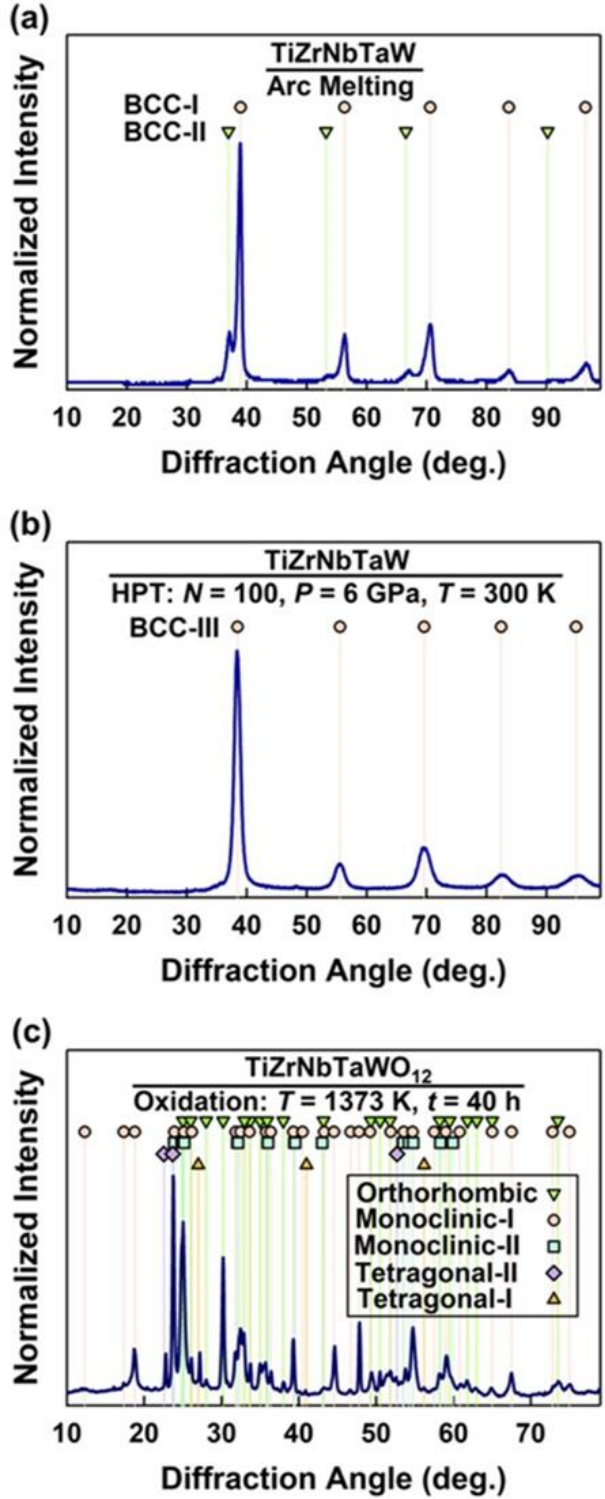

Fig. 2 Presence of one orthorhombic, two monoclinic and two tetragonal phases in HEO. XRD profiles for equiatomic TiZrNbTaW alloy after (a) arc melting, (b) HPT processing, and (c) oxidation.

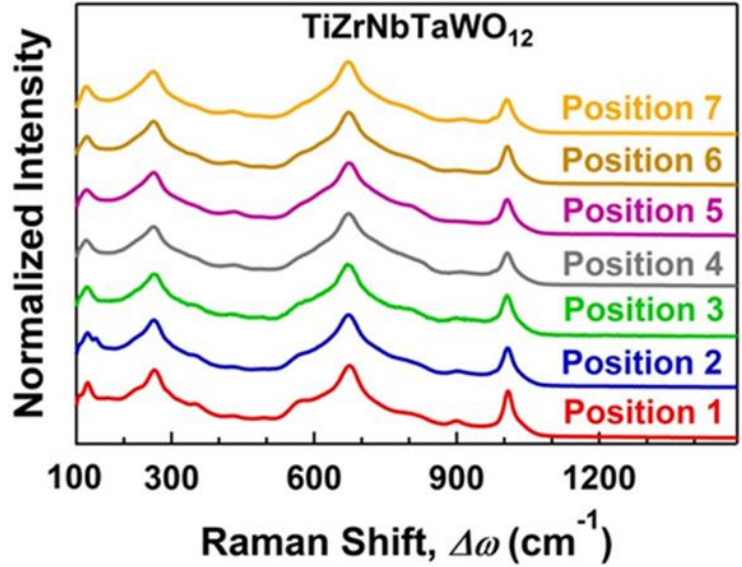

Fig. 3 Raman spectroscopy of HEO taken at seven different positions of synthesized sample.

Table 1 Crystallographic information, phase fractions, particle size, grain size and specific surface area of HEO examined by XRD, Rietveld refinement, SEM, DLS, TEM and BET methods.

| Characteristics | Method | Orthorhombic | Monoclinic-I | Monoclinic-II | Tetragonal-I | Tetragonal-II |
|---|---|---|---|---|---|---|
| Space Group | XRD | Ima2 | C12/m1 | P12/m1 | P4/nmm | P42/mnm |
| Lattice parameters (nm) | Rietveld | $a = 4.099$<br>$b = 0.492$<br>$c = 0.529$ | $a = 2.037$<br>$b = 0.379$<br>$c = 1.543$ | $a = 1.934$<br>$b = 0.380$<br>$c = 1.994$ | $a = b = 0.5291$<br>$c = 0.394$ | $a = b = 0.460$<br>$c = 0.297$ |
| Lattice Angles | Rietveld | $\alpha = \beta = \gamma = 90°$ | $\alpha = \gamma = 90°$<br>$\beta = 113.4$ | $\alpha = \gamma = 90°$<br>$\beta = 115.7$ | $\alpha = \beta = \gamma = 90°$ | $\alpha = \beta = \gamma = 90°$ |
| Phase Fraction (wt%) | Rietveld | 29.4 | 38.6 | 26.4 | 3.9 | 1.7 |
| Visualized Structure | VESTA | 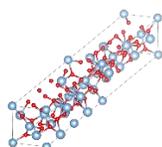 | 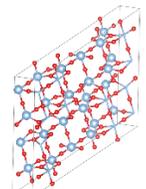 | 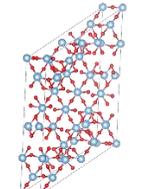 | 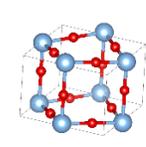 | 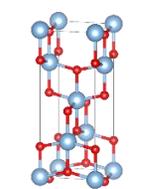 |
| Particle Size (nm) | SEM<br>DLS | | | 470<br>408 | | |
| Grain Size (nm) | TEM | | | 110 | | |
| Surface Aera (m²g⁻¹) | BET | | | 0.76 | | |

The distribution of elements at the micrometer and nanometer scales are shown in Fig. 4(a) and (b) using the SEM-EDS and STEM-EDS mappings, respectively. The SEM-EDS analysis suggests that the general composition of materials is TiZrNbTaWO$_{12}$, which is consistent with the measured mass rise of the sample after oxidation. The EDS elemental mappings show that despite heterogeneities in the distribution of elements and segregation of Ti in some regions, the five metallic elements and oxygen are largely available in all areas, as shown in the EDS point analyses in Table 2. This suggests that the five phases are not in the form of binary or ternary oxides, and they can have medium- or high-entropy features, although it is hard to determine their exact compositions using EDS [28-30]. Here it should be noted that the XRD analysis and the Rietveld refinement also did not provide clear evidence for the presence of binary or ternary oxides.

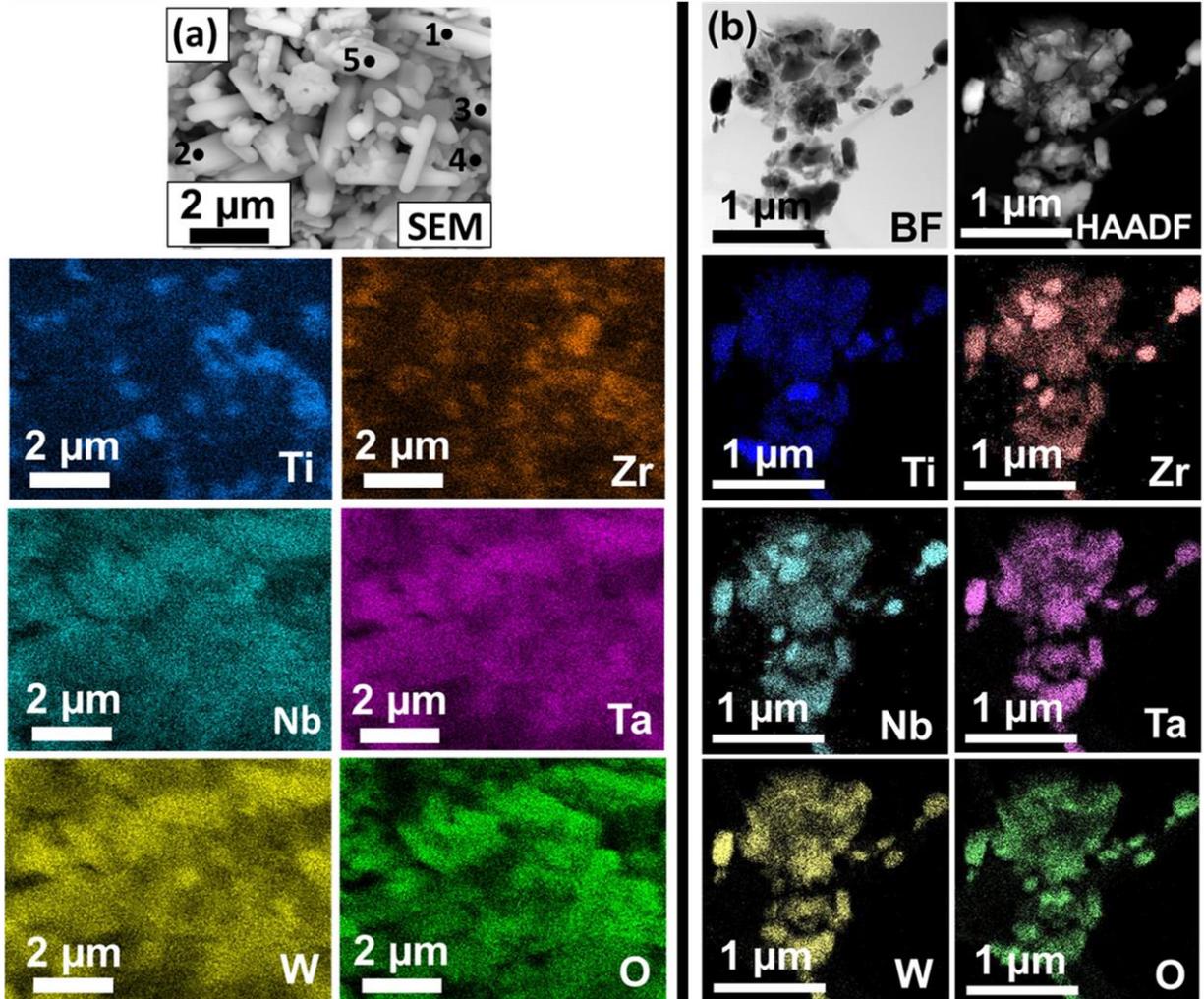

Fig. 4 Distribution of elements in HEO. (a) SEM and (b) STEM-HAADF images and corresponding elemental mappings using EDS analysis.

Table 2. Quantitative EDS analysis for fraction of elements (in at%) in HEO for points indicated in SEM image of Fig. 4a.

| Point | Ti  | Zr  | Nb  | Ta   | W    | O    |
|-------|-----|-----|-----|------|------|------|
| 1     | 3.3 | 2.3 | 6.4 | 5.7  | 5.3  | 77.0 |
| 2     | 4.2 | 2.9 | 7.9 | 7.9  | 7.0  | 70.1 |
| 3     | 7.0 | 4.2 | 8.3 | 10.4 | 11.8 | 58.3 |
| 4     | 9.1 | 9.8 | 4.2 | 6.1  | 3.8  | 67.0 |
| 5     | 5.1 | 1.8 | 5.9 | 5.5  | 3.5  | 78.2 |

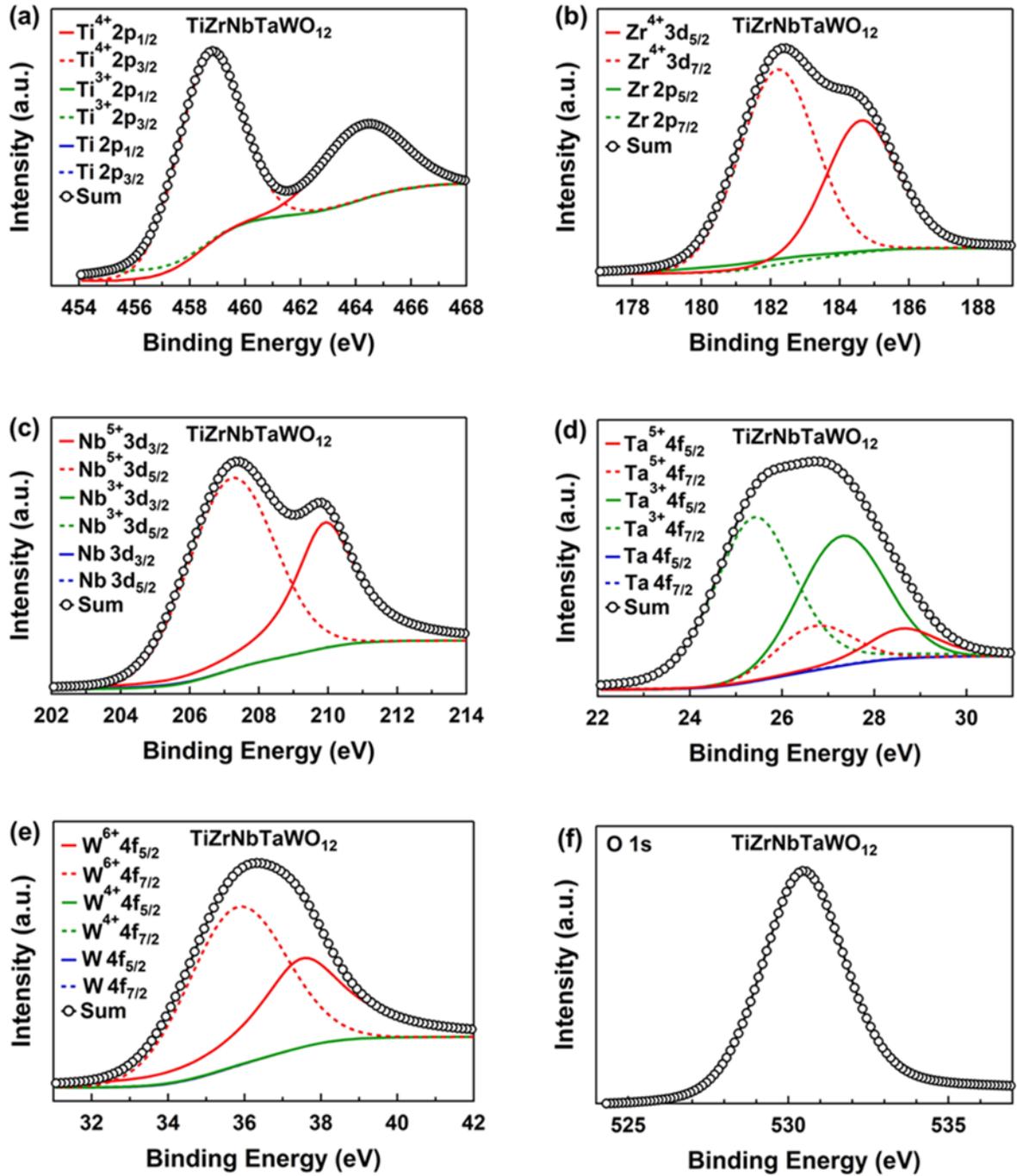

Fig. 5 Oxidation and reduction states of cations and oxygen anion in HEO. XPS spectra of (a) Ti 2p, (b) Zr 3d, (c) Nb 3d, (d) Ta 4f, (e) W 4f, and (f) O 1s and corresponding peak deconvolutions.

Examination of the oxidation state of elements using XPS is shown in Fig. 5 for (a) Ti, (b) Zr, (c) Nb, (d) Ta, (e) W and (f) O. All elements are mainly in the fully oxidized states (i.e., $Ti^{4+}$, $Zr^{4+}$, $Nb^{5+}$, $Ta^{5+}$ and $W^{6+}$), although some lower oxidation states are detected by the peak deconvolution method. These XPS analyses suggest that primary metallic alloy could be successfully converted to oxide with the $d^0$ electronic configuration. These results confirm the potential of HPT processing followed by high-temperature oxidation as an effective route for the synthesis of HEOs, as mentioned in earlier publications [51,52].

*3.2. Microstructure*

The microstructure of oxide is shown in Figs. 6(a), (b) and (c) using the BF image, SAED analysis and DF image, respectively. The DF image was taken by the diffracted beams indicated by an arrow in the SAED analysis. The presence of ultrafine grains can be observed in the DF image, in which many white regions with nanometer and submicrometer sizes exist. The presence of ultrafine grains can be also confirmed from the SAED analysis with an approximate ring shape. Fig. 7 shows the grain and particle size distribution for the oxide using (a) TEM, (b) SEM and (c) DLS analyses. The grain sizes are in the range of several nanometers to ~300 nm with an average size of 110 nm. Fig. 7(b) shows that the particle sizes measured by SEM vary in a wide range from the nanometer level to ~1400 nm. Examination of particle size distribution by DLS in Fig. 7(c) also shows that the particles have a wide range of sizes from the nanometer level to ~800 nm. Despite the differences in the distribution of particle sizes in Figs. 7(b) and (c), the average particle sizes measured using the two methods are reasonably similar: 470 nm using SEM and 408 nm using DLS.

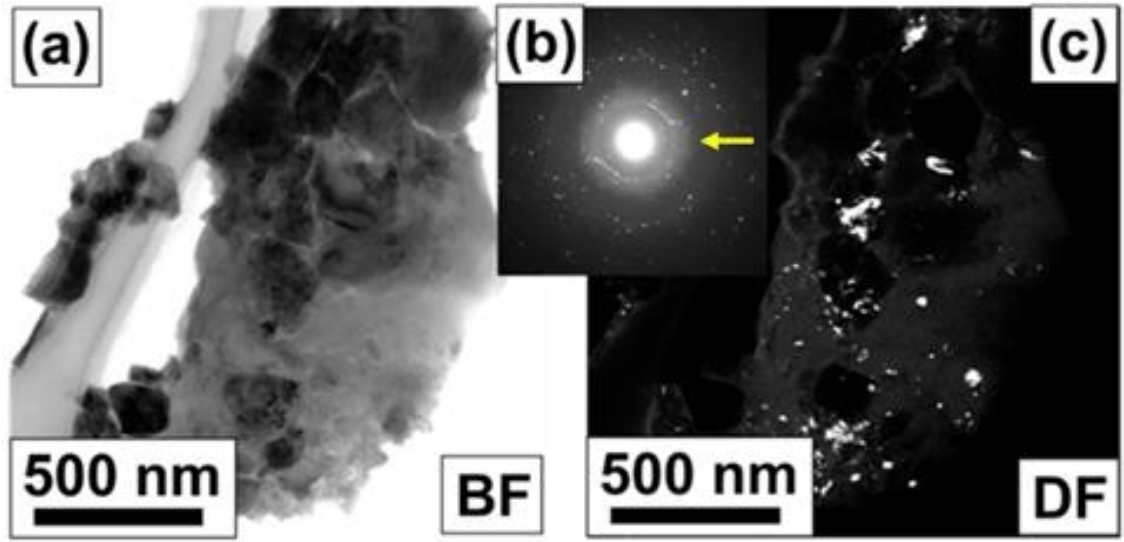

Fig. 6 Presence of nanocrystals in HEO. TEM (a) BF, (b) SAED. and (c) DF images.

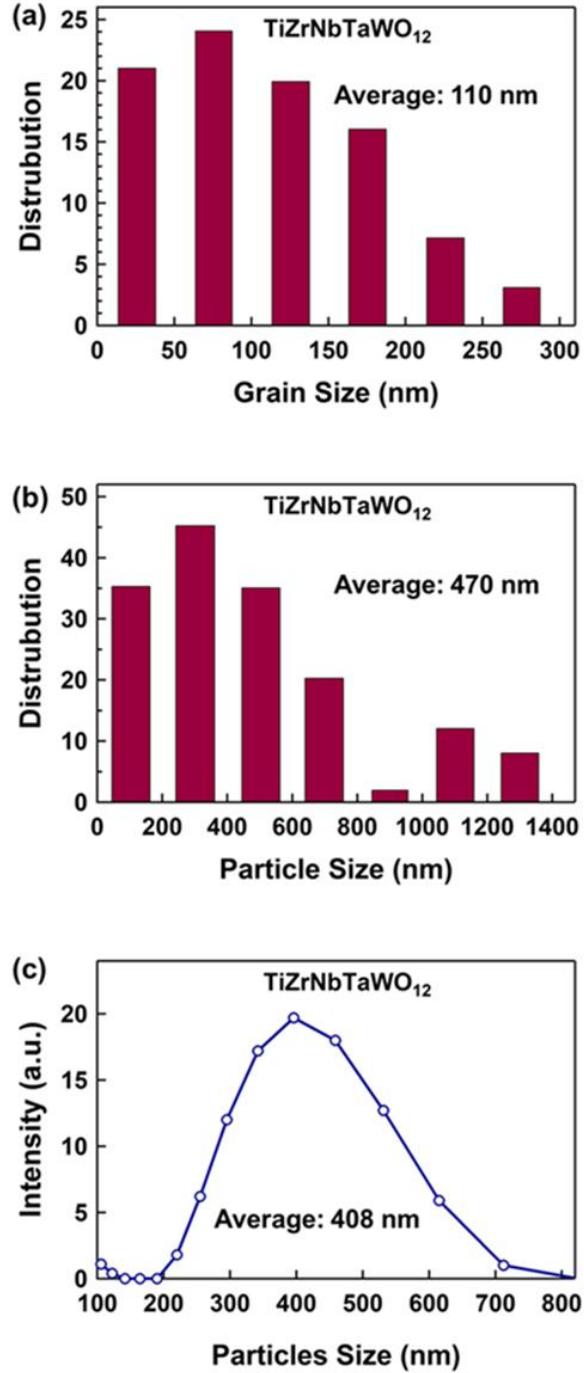

Fig. 7 Grain size and particle size distribution in HEO. (a) Grain size distribution measured by TEM analysis, and particle size distribution measured by (b) SEM and (c) DLS analysis.

Examination of nanostructure using high-resolution TEM images and FFT diffractograms, as shown in Fig. 8, also confirms the presence of five phases, in good agreement with the XRD analysis: (a) orthorhombic, (b) monoclinic-I, (c) monoclinic-II, (d) tetragonal-I and (e) tetragonal-II. Here, it should be noted that the formation of ultrafine-grained phases is a general feature of HPT-processed or HPT-synthesized materials [50,52]. The formation of heterojunctions between phases is shown in Fig. 9 using TEM high-resolution images. Totally 9 out of 10 possible heterojunctions could be detected. The only heterojunction that could not be found in about 30 high-resolution images was the boundary between the monoclinic-II and tetragonal-II phases.

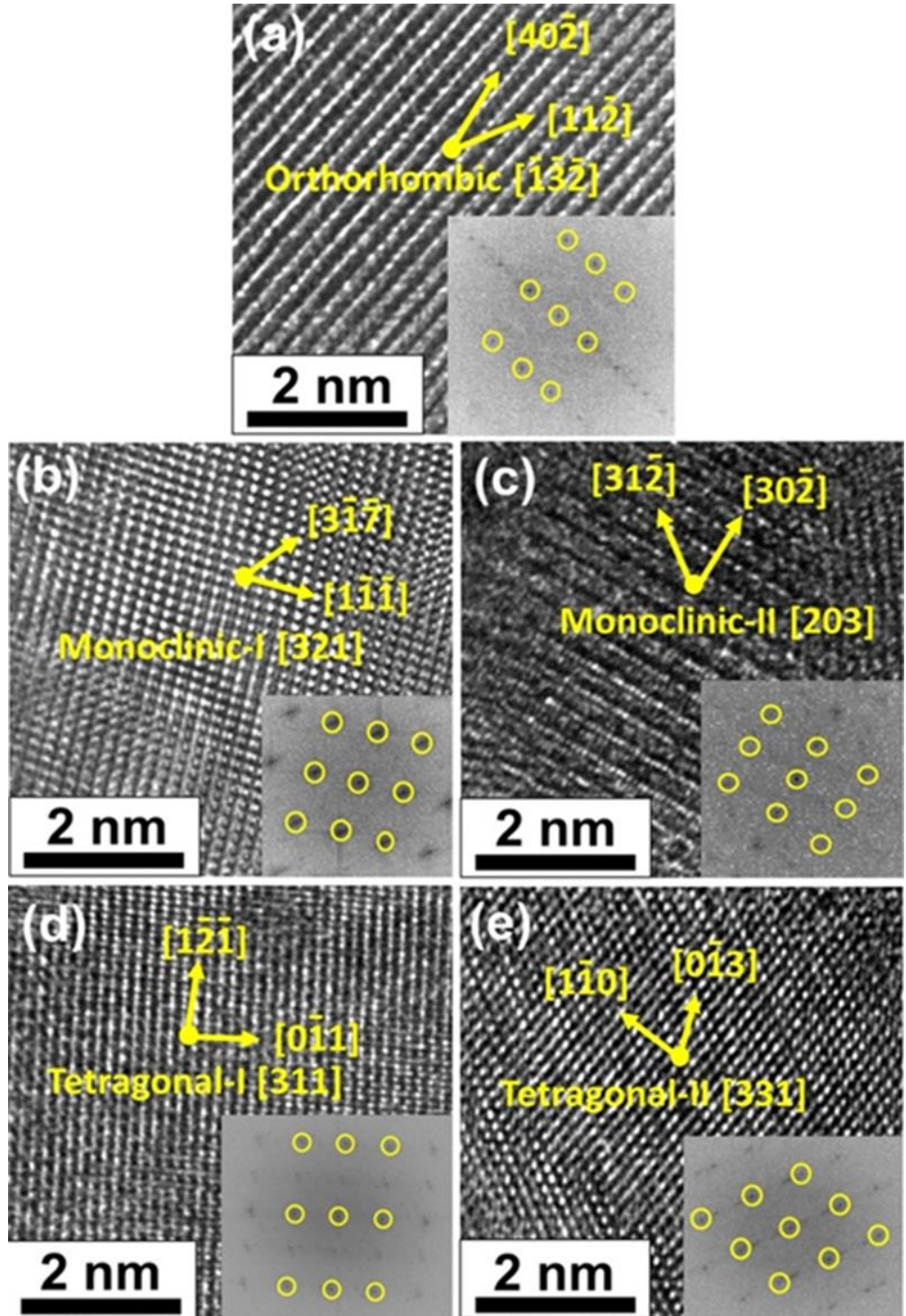

Fig. 8 Presence of five phases in HEO. TEM lattice images and corresponding FFT diffractogram for (a) orthorhombic (Ima2), (b) monoclinic-I (C12/m1), (c) monoclinic-II (P12/m1), (d) tetragonal-I (P4/nmm), and (e) tetragonal-II (P42/mnm) phases.

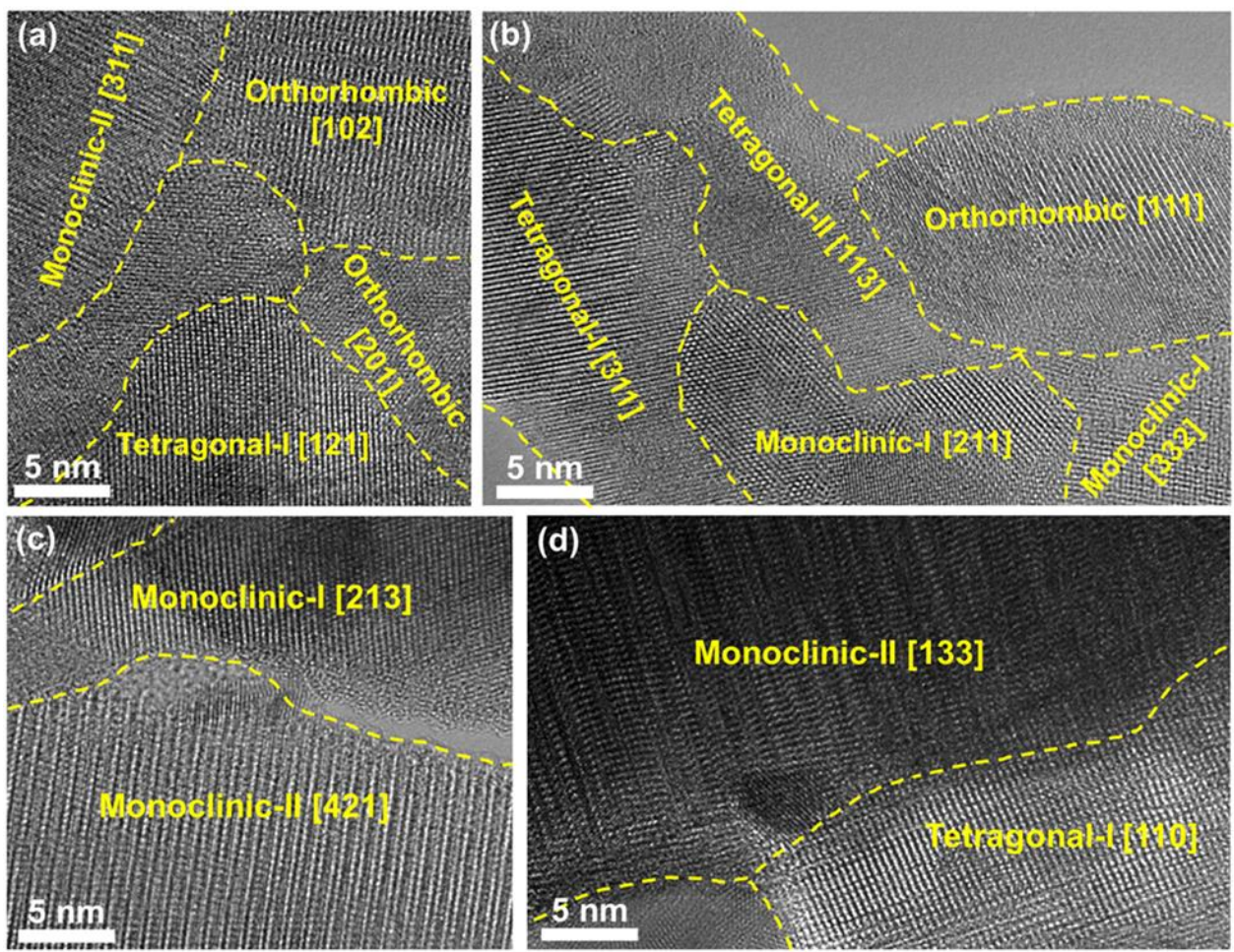

Fig. 9 Formation of numerous heterojunctions in HEO. TEM high-resolution images showing heterojunctions between (a) orthorhombic/monoclinic-II, orthorhombic/tetragonal-I, (b) orthorhombic/monoclinic-I, orthorhombic/tetragonal-II, monoclinic-I/tetragonal-I, monoclinic-I/tetragonal-II, tetragonal-I/tetragonal-II, (c) monoclinic-I/monoclinic-II and (d) monoclinic-II/tetragonal-I.

### 3.3. Electronic structure

The electronic structure of oxide which is investigated by UV-vis and UPS is presented in Fig. 10, where (a) shows the light absorbance of the material, (b) shows the Kubelka-Munk plot to estimate the indirect bandgap ($E_g$), (c) shows the UPS spectrum to calculate the top of the valence band and (d) shows the electronic band structure by considering both UV-vis and UPS spectra. Fig. 10(a) shows that the oxide can absorb light in both UV and visible regions, although the amount of light absorbance in the UV region is naturally more significant. Fig. 10(b) shows the presence of two bandgaps with energy levels of 2.8 and 2.3 eV. Although it is hard to mention which these bandgaps belong to which phase(s), it is evident that both bandgaps are in the visible-light region of sunlight. From the UPS spectrum of Fig. 10(c), it can be shown that the cut-off energy from the Fermi level ($E_0$) and the valence band top energy from the Fermi level ($E_B$) are 15.4 and 0.8 eV, respectively, and thus, the top of the valence band from the vacuum level is calculated as $E_{VBT} = -21.2 + (E_0 - E_B) = -6.6$ eV. If the bottom of the conduction band ($E_{CBM}$) is calculated as $E_{CBM} =$ $E_{VBT} + E_g$ = -3.8 eV and -4.4 eV, the electronic band structure of oxide can be summarized in Fig. 10(d). This estimated electronic band structure illustrates that both positions of the bottom of the conduction band are higher than the chemical potential for the reduction of water to $H_2$ and the position of the top of the valence band is lower than the chemical potential for the oxidation of water to $O_2$. This band structure suggests that the oxide can basically act as a photocatalyst for water splitting under visible light [1-10]. Although first-principle electronic structure calculations are needed to clarify the origin of visible-light absorbance in this oxide, the presence of W is critical in bandgap narrowing because the light absorbance is limited to the UV light when W is replaced with another element such as Hf [25,27].

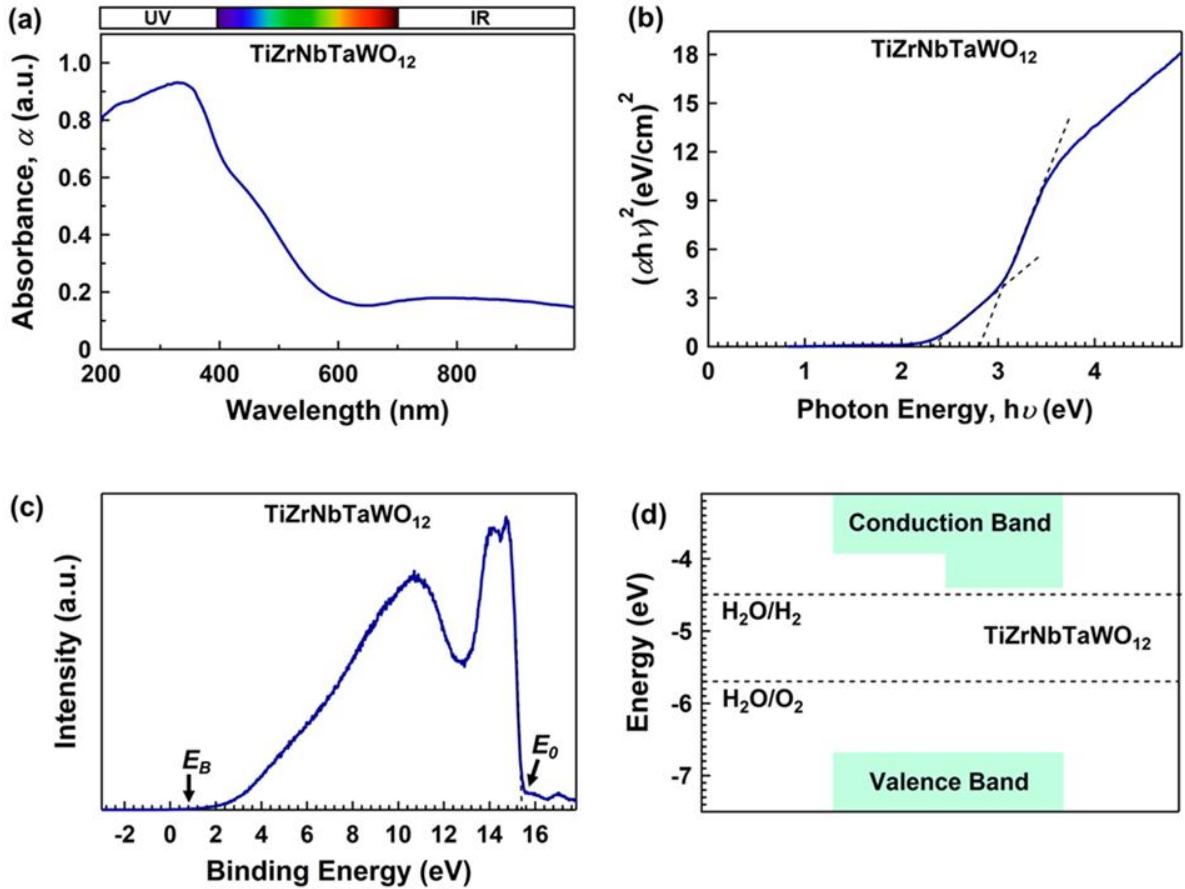

Fig. 10 Appropriate light absorbance and band structure of HEO for photocatalytic water splitting. (a) UV-vis light absorbance spectrum, (b) Kubelka-Munk plot for indirect bandgap calculation (α: light absorption, h: Planck's constant, ν: light frequency), (c) UPS spectrum for top of valence band calculation, and (d) electronic band structure compared with standard chemical potentials for water splitting reactions.

Examination of electron-hole recombination using steady-state PL spectroscopy is shown in Fig. 11. There is a PL peak at a wavelength of 570 nm or at an energy level of 2.2 eV. Since this energy level is close to the energy gap of 2.3 eV observed in Fig. 10(b), it is concluded that electron-hole recombination occurs at this low energy gap [55-58]. However, the absence of a clear peak at a wavelength of 440 nm (there is a very weak shoulder) confirms that the recombination rate is less for the bandgap of 2.8 eV, suggesting that the excited electron within this bandgap can have enough time for distribution in the heterojunctions [17-24]. Moreover, there is a possibility that the photoluminescence emitted at 440 nm is reabsorbed by the material [55].

It should be noted that even the PL intensity for the peak at 570 nm (880 cps) is not so significant for a photocatalyst because anatase $TiO_2$ photocatalyst shows a PL peak with one order of magnitude higher intensity (12,300 cps) under similar PL testing conditions [27]. Moreover, many strategies developed for visible-light-driven photocatalysis can also lead to similar PL intensities due to defect-induced or impurity-related recombination effects [11-16]. It is then concluded that the electron-hole recombination in this oxide is in a reasonable range when compared to other photocatalysts.

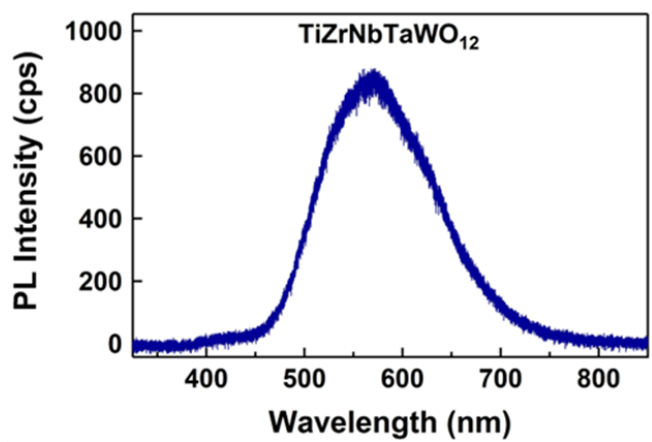

Fig. 11 Steady-state PL emission of HEO achieved using 325 nm laser irradiation.

*3.4. Photocatalytic activity*

The photocatalytic activity of oxide for water splitting under visible light is shown in Fig. 12(a). Although the specific surface area of this oxide is rather small for a catalyst (0.76 $m^2g^{-1}$ measured by BET as given in Table. 1), it successfully produces oxygen from water under visible light without co-catalyst addition. In the presence of a sacrificial agent, the oxide generates 1.4 µmol of oxygen after 300 min, while the standard deviations for three different measurements were less than 10%. This indicates that holes effectively take part in the oxidation reaction of water [1,5]. Despite the appropriate band structure of oxide for the water reduction reaction, no hydrogen production is detected under visible light with or without the addition of a sacrificial agent within the detection limits of gas chromatography. The blank tests confirmed that no oxygen is produced without irradiation and with catalyst addition (datum at the time of zero) as well as with irradiation and without catalyst addition, indicated as blank in Fig. 12(a). Examination of the oxide after the photocatalytic test using XRD analysis, as shown in Fig. 12(b), confirms that the structure of the material is quite stable during photocatalysis. It should be noted that slight changes in the XRD profiles at 15-20º and 30-40º are due to the small amount of recovered material after photocatalysis which results in the appearance of a background effect from the sample holder, which has two humps at 15-20º and 30-40º. The stability of this catalyst particularly under visible light is not surprising because high-entropy materials usually show high stability due to their low Gibbs free energy, as shown in many other applications including Li-ion batteries [31-33], magnetic components [34-36], dielectric components [37-39] and thermomechanical applications [40-42]. In addition to the investigation of crystal structure stability by XRD, examination of the surface stability by the XPS analysis after the photocatalytic test, as shown in Fig. 13, demonstrates no significant changes compared to Fig. 5. Moreover, UV-vis spectra of this oxide before and after photocatalysis were confirmed to be quite similar. The TEM analysis of the oxide also did not show any clear difference before and after the photocatalysis. Taken altogether, the current oxide shows activity for photocatalysis under visible light, but future studies are required to determine the wavelength dependence of this activity.

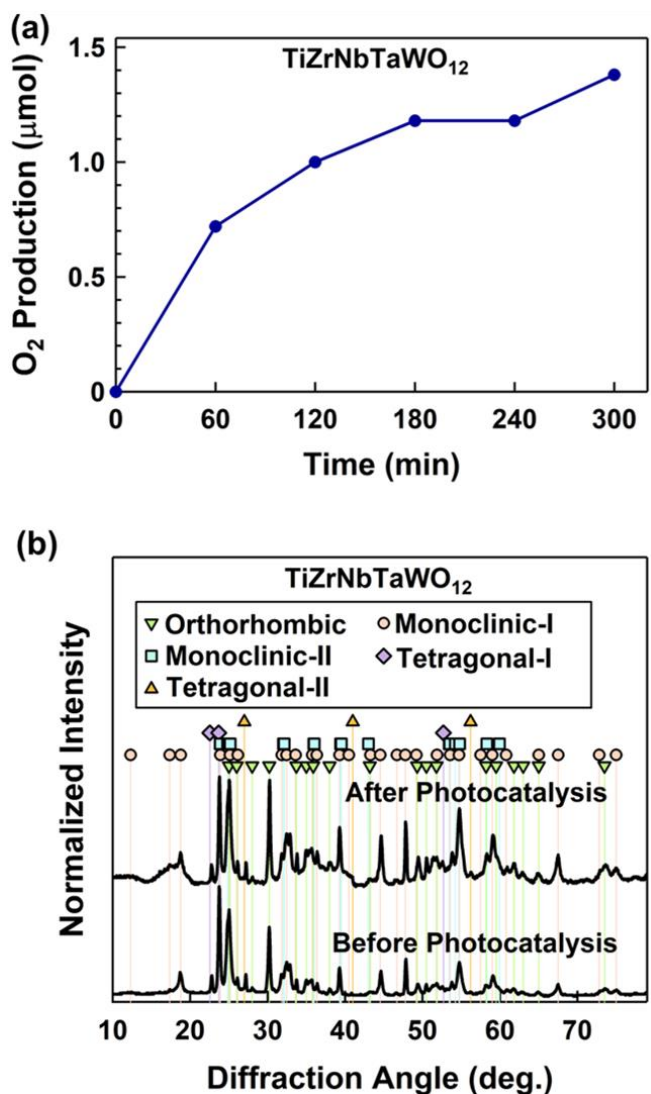

Fig. 12 Photocatalytic oxygen evolution and structural stability of HEO. (a) Oxygen production amount under visible light versus irradiation time and (b) XRD profiles of oxide before and after photocatalytic test.

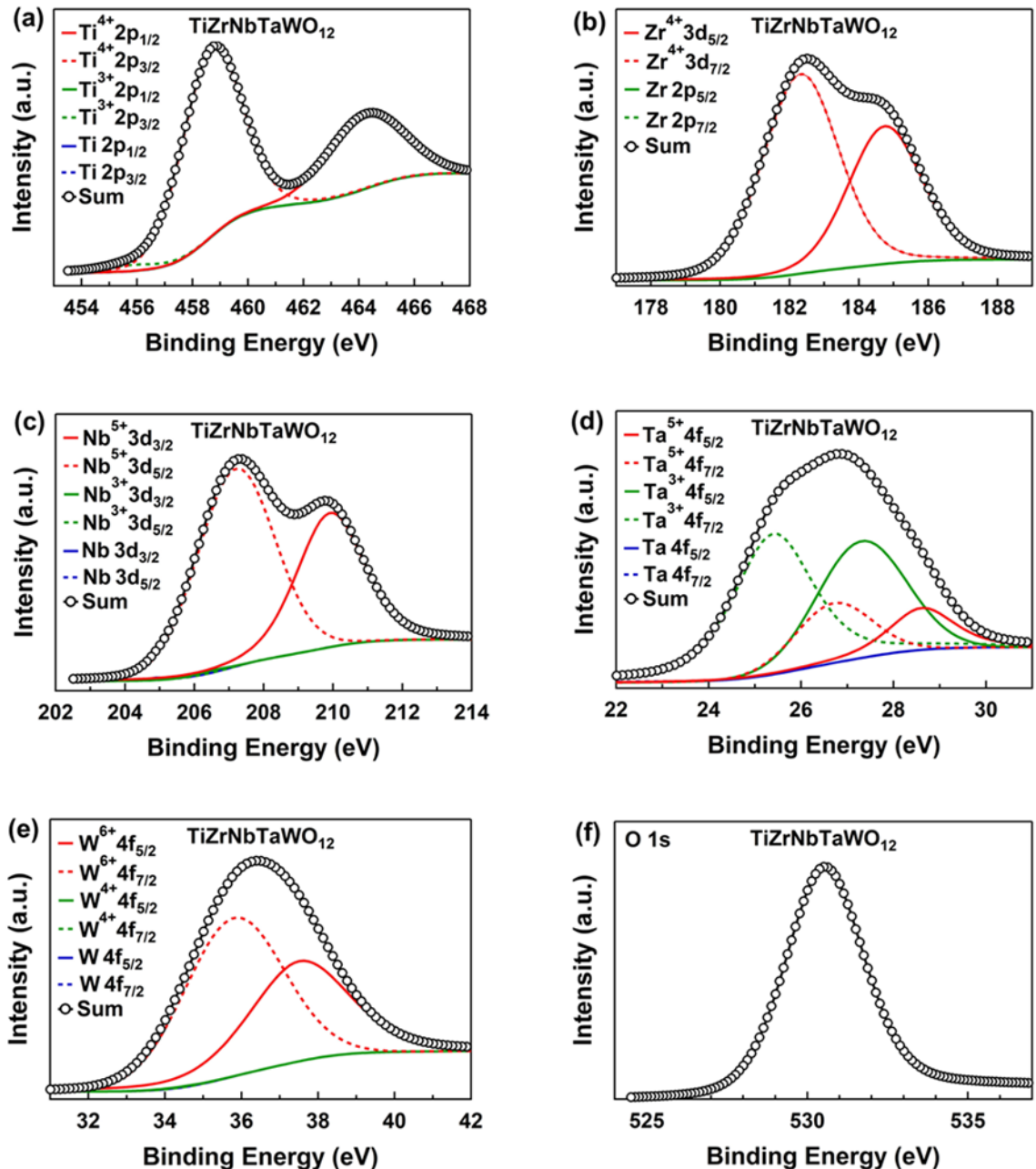

Fig. 13 Surface stability of HEO after photocatalytic test. XPS spectra of (a) Ti 2p, (b) Zr 3d, (c) Nb 3d, (d) Ta 4f, (e) W 4f, and (f) O 1s and corresponding peak deconvolutions.

## 4. Discussion

    Here, a concept of high-entropy materials is combined with the concept of heterojunctions to produce a new visible-light-active photocatalyst. Since there is still quite limited information about the band structure of high-entropy photocatalysts, it is hard to select the best pairs to generate an efficient heterojunction for photocatalysis. To overcome this barrier, 10 different heterojunctions were introduced into the oxide in this study. Multiple heterojunctions were introduced because of two main reasons. First, a few earlier studies suggested that using three

components/phases for heterojunction improves the charge carrier separation and photocatalytic activity compared to the introduction of heterojunctions in dual-phase or two-component systems [18,59,60]. Second, the possibility of having one appropriate heterojunction for visible-light-driven photocatalysis statistically increases with increasing the number of heterojunctions [25-27]. The strategy of multiple-heterojunction introduction appears as a successful solution to achieving visible-light-driven photocatalytic oxygen production. In addition to the effect of multiple heterojunctions on charge carrier separation [17,18], the high-entropy concept is also effective in higher light absorbance and high stability of photocatalysts, as reported in HEOs [28-30].

Here, it is worth comparing the current photocatalyst with the binary oxides in the Ti-Zr-Nb-Ta-W-O system. As shown in Fig. 14, the light absorbance of current high-entropy oxide is higher than the light absorbance of binary oxides $TiO_2$, $ZrO_2$, $Nb_2O_5$, $Ta_2O_5$ and $WO_3$. Table 3 compares the photocatalytic activity of current HEO with binary oxides for oxygen production under visible light and with the addition of $AgNO_3$ as a sacrificial agent. $TiO_2$, $ZrO_2$, $Nb_2O_5$ and $Ta_2O_5$ with wide bandgap do not produce any oxygen and $WO_3$, which has a low bandgap, generates 7.9 µmolh$^{-1}$m$^{-1}$ $O_2$. The amount of oxygen production for the high-entropy oxide is 12.1 µmolh$^{-1}$g$^{-1}$ which is higher than $WO_3$, although the specific surface area of this high-entropy oxide (0.76 m$^2$g$^{-1}$) is lower compared to $WO_3$ (2.72 m$^2$g$^{-1}$). Despite the large light absorbance and good visible-light activity of this material, enhancing its surface area by using other synthesis methods, developed for HEOs [28-47], is necessary to maximize its photocatalytic activity. Surface area is an important factor in photocatalytic activity. For example, $WO_3$ nanosheets with a large surface area can show higher activity under visible light compared to the commercial $WO_3$ used in this study [61]. Although, various parameters such as light source, catalyst mass, pH and photoreactor type influence the photocatalytic performance, but a comparison between the oxygen production rate of current high-entropy oxide with other reported photocatalysts can still give an overall insight regarding the efficiency of this high-entropy oxide. Table 4 [62-80] makes such a comparison based on the oxygen production amount per catalyst mass and per catalyst surface area. Since the photocatalytic reaction occurs on the catalyst surface, the production amount per surface area may provide a more reliable comparison. Oxygen production in Table 4 varies in the range of 0.002-38.75 µmolh$^{-1}$m$^{-2}$, and the production amount of 12.1 µmolh$^{-1}$m$^{-2}$ on the high-entropy oxide is reasonably high among the reported values.

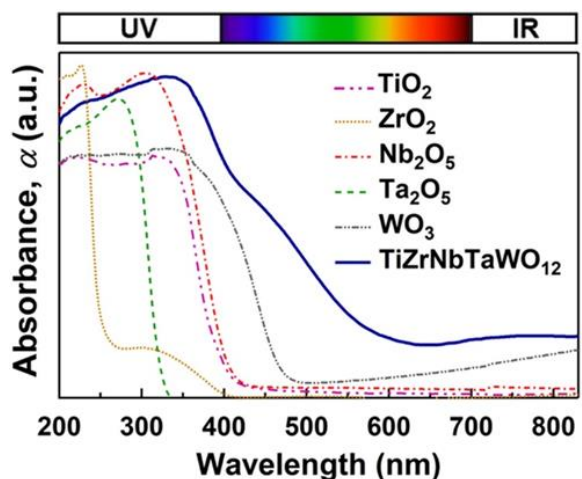

Fig. 14 High light absorbance of HEO compared with relevant binary oxides $TiO_2$, $ZrO_2$, $Nb_2O_5$, $Ta_2O_5$ and $WO_3$.

Table 3 Amount of oxygen production under visible light and bandgap of HEO compared to relevant binary oxides.

| Photocatalyst | Bandgap (eV) | $O_2$ production ($\mu mol h^{-1} m^{-2}$) |
|---|---|---|
| $TiO_2$ | 3.0 | 0 |
| $ZrO_2$ | 5.1 | 0 |
| $Nb_2O_5$ | 3.2 | 0 |
| $Ta_2O_5$ | 3.9 | 0 |
| $WO_3$ | 2.5 | 7.9 |
| $TiZrNbTaWO_{12}$ | 2.3 & 2.8 | 12.1 |

Table 4 Amount of oxygen production under visible light on HEO compared to other photocatalysts reported in literature.

| Photocatalyst | Catalyst Mass (mg) | Light Source Before Cut-Off Filter | $O_2$ Production ($\mu mol h^{-1} g^{-1}$) | $O_2$ Production ($\mu mol h^{-1} m^{-2}$) | Reference |
|---|---|---|---|---|---|
| $WO_3$ / Red Phosphorous | 50 | 300 W Xe | 2.7 | 0.13 | [62] |
| $CoO_x$ / $ScTaO_{4-x}N_x$ | 100 | 300 W Xe | ~36.7 | --- | [63] |
| Co-Doped ZnO Nanorod | 50 | 300 W Xe | 191 | 33.5 | [64] |
| La/Rh-Doped $SrTiO_3$ | 100 | 300 W Xe | ~220 | 26.73 | [65] |
| Carbon Nitride Sheets | 50 | 300 W Xe | 22.6 | 0.11 | [66] |
| Ag / $ZnIn_2S$ | 5 | 300 W Xe | 29.1 | 0.67 | [67] |
| BiOBr / C | --- | 150 W Xe | 110 | - | [68] |
| $TiO_2$ / BiOBr | 50 | 300 W Xe | 95.7 | 0.96 | [69] |
| Few-Layer $BiVO_4$ Nanosheet | 70 | 300 W Xe | 63.7 | 1.26 | [70] |
| Zr/Ce/Ti Metal Organic Framework | 20 | UV-Vis xenon | 2.5 | 0.0025 | [71] |
| $WO_3 \cdot H_2O$ / $g-C_3N_4$ | 100 | 300 W Xe | 232 | 1.46 | [72] |
| $RuO_2$ / $Sr_2CoTaO_6$-F | 100 | 300 W Xe | 60 | 18.64 | [73] |
| $RuO_2$ / $Sr_2CoTaO_6$-S | 100 | 300 W Xe | 30 | 38.75 | [73] |
| Rh / $Sr_2CoTaO_6$-F | 100 | 300 W Xe | 4 | 1.24 | [73] |
| Rh / $Sr_2CoTaO_6$-S | 100 | 300 W Xe | 1 | 1.3 | [73] |
| $TiO_2$ Nanobelts | 20 | 500 W Xe | 410 | 4.31 | [74] |
| Pt-$TiO_2$ Nano-Architecture | 10 | 300 W Xe | 156.2 | 1.04 | [75] |
| La/N-Doped $Sr_2TiO_4$ | 100 | 300 W Xe | 48 | 37.64 | [76] |
| Covalent Triazine Frameworks | 100 | 300 W Xe | 36.7 | 0.18 | [77] |
| $Co_3(PO_4)_2$ / $g-C_3N_4$ | 50 | 300W Xe | 177.3 | 1.75 | [78] |
| CoO/$g-C_3N_4$ | 50 | white Light Diode | 27.8 | 0.85 | [79] |
| NiO / Carbon Dots / $BiVO_4$ | 10 | 300 W Xe | 60 | - | [80] |
| $TiZrNbTaWO_{12}$ | 30 | 300 W Xe | 9.2 | 12.1 | This study |

Although earlier studies showed the catalytic activity of high-entropy materials for $H_2$ production and $CO_2$ conversion [25-27] the current study introduces high-entropy oxides as new photocatalysts for visible-light-driven $O_2$ production. Moreover, this study suggests the significance of multiple heterojunctions in tailoring the photocatalytic activity to the visible-light region of sunlight. Combining this strategy with the existing knowledge on heterojunctions in two-phase/component and three-phase/component materials [18-24,59,60] is expected to lead to the production of various active photocatalysts for different applications.

## 5. Conclusions

The first application of high-entropy oxides as photocatalysts for oxygen production under visible light was reported in this study. A new photocatalyst in the Ti-Zr-Nb-Ta-W-O system was fabricated by arc melting, followed by high-pressure homogenization and high-temperature oxidation. It had 10 types of heterojunctions and could act as a photocatalyst for oxygen production under visible light without co-catalyst addition. This activity is due to large light absorbance, low bandgap and appropriate electronic band structure which are due to the presence of multiple phases and heterojunctions in this material. This work can open a new pathway to design new photocatalysts with high activity for water splitting under visible light.

**Declaration of Competing Interest**

There are no conflicts to declare.

**Acknowledgements**

This work is supported in part by the WPI-I2CNER, Japan, and in part by Grants-in-Aid for Scientific Research on Innovative Areas from the MEXT, Japan (JP19H05176 & JP21H00150).